\documentclass[aps,prl,twocolumn,superscriptaddress,showpacs]{revtex4}
\usepackage{amssymb}
\usepackage{graphicx}
\usepackage{amsmath}
\usepackage{amsfonts}

\begin{document}

\title{Superconductivity in a molecular graphene}

\author{Jin-Hua Gao}
\affiliation{School of Physics and Wuhan National High Magnetic field center,
Huazhong University of Science and Technology, Wuhan 430074,  China }

\author{Yi Zhou}
\affiliation{Department of Physics, Zhejiang University, Hangzhou 310027,  China}
\affiliation{Collaborative Innovation Center of Advanced Microstructures, Nanjing University, Nanjing 210093, China}

\author{Fu-Chun Zhang}
\email{fuchun@hku.hk}
\affiliation{Department of Physics, Zhejiang University, Hangzhou 310027, China}
\affiliation{Collaborative Innovation Center of Advanced Microstructures, Nanjing University,  Nanjing 210093, China}

\begin{abstract}
We propose that constructing a molecule super-lattice on a superconducting ultrathin film is a promising way to manipulate superconductivity in experiment.
We theoretically study superconductivity in a molecule graphene system, which is built by fabricating a hexagonal molecule super-lattice  on  2-dimensional electron gas.
The super-lattice potential dramatically changes the electron density of states, which oscillates as function of the energy.
We show that such a molecular graphene may increase superconducting gap by a few times, which may open a new route to realize high temperature superconductivity.
\end{abstract}

\maketitle

Superconductivity is a novel phenomenon, in which electric resistance vanishes below a critical temperature.
There have been great activities in recent decades in discovering new classes of superconductors with high transition temperatures.
The known techniques to manipulate superconductivity include applications of pressure, electric gating, as well as chemical doping.
The latter one has so far the most successful. The undoped cuprates such as La$_2$CuO$_4$ are Mott insulator,
and high temperature superconductivity occurs when charge carriers are introduced by chemical substitutions\cite{muller,anderson}.
Chemical doping also plays a key role in more recently discovered iron-based high temperature superconductors\cite{hosono,NSR}.
Meanwhile, quantum confinement is also a possible way in low dimensional system\cite{thompson,1dsc,2dsc}.
In this Letter, according to  recent  experimental advances,
we propose a new promising route to manipulate superconductivity by constructing an artificial super-lattice  on a quasi two-dimensional (2D) superconductor
via molecule absorption. We study the superconductivity in a molecule graphene system, which is built by fabricating
a molecule  hexagonal super-lattice on a two-dimensional electron system (2DES)\cite{manoharan2012,slouie2009}.
Such artificial super-lattice dramatically changes the electronic structure and may greatly increase density of states (DOS) at the Fermi level,
hence to substantially increase the transition temperature.

 Similar molecule super-lattices, as well as the molecule graphene system, have experimentally been realized in 2DES  on the surface state of copper crystal recently\cite{manoharan2012}.
 In this case, hundreds of carbon monoxide molecules absorbed on copper crystal surface are assembled into a hexgonal lattice
 by positioning the molecules individually with the tip of a scanning tunneling microscope (STM).
 The carbon monoxide molecules act as a repulsive potential, i.e the lateral super-lattice,
 and effectively confine the motion of surface electrons to a honeycomb lattice, leading to an artificial Dirac fermion system in analogy to graphene\cite{manoharan2012,slouie2009,manoharan2008}.
 The linear dispersion and the DOS of the Dirac fermion have been observed in the STM measurements.
 Here we point out that such a molecule super-lattice potential applied to a superconducting (SC) ultrathin film (quasi 2D superconductor) provides a new platform
 to study the properties of superconductivity and to raise the SC transition temperature.
 It is interesting to note that, in experiment, the single electron states in a 2DES with a quadratic dispersion have been changed to massless Dirac fermions
 with linear dispersion near certain symmetry points by applying suitable molecule super-lattice potential.
 The DOS in a conventional 2DES is a constant, while the DOS in such an artificial Dirac fermion system is a function of energy
 and can be tuned by designing the lattice constant as well as the potential of the super-lattice, as have been demonstrated in experiments.
 Therefore, such molecule super-lattices may provide an effective and feasible way to manipulate the superconductivity.

The embedded super-lattice potential technique is only applicable to 2D electron system. The 2D SC state is generally believed to be a fragile state of matter,
which does not have true long range order at a finite temperature\cite{np2004}. Instead, the system may undergo a Kosterlitz-Thouless transition at
a finite temperature $T_{\textrm{BKT}}$, below which the SC order parameter decays in space in a power law.
Note that recent experiments on ultrathin metal film have indicated that superconductivity persists even in films with 1-2 atomic layer
thickness\cite{yguo,shih2006,shih2009,tzhang2010,nakayama2011,okamoto2013}. To be concrete, in this paper we study the zero temperature SC gap $\Delta_0$
in the artificial super-lattice, while leaving the nature of the finite temperature transition for future discussion.
In a conventional superconductor, $\Delta_0 \varpropto \omega_D e^{-1/N(0)V}$, with $N(0)$ the DOS at the Fermi level, $V$ the electron-phonon coupling,
and $\omega_D$ the Debye cutoff energy.  We may tune $N(0)$ hence $\Delta_0$ easily, thanks to the artificial super-lattices.
Our theoretical study should  stimulate further experiment activities of superconductivity in 2D systems.

To make our proposal more concrete and quantitative, in what follows we will consider 2DES under a hexagonal super-lattice potential
and calculate the electronic structure.  The DOS in such a molecular  graphene is found to strongly oscillate and to depend on the Fermi level
at the lowest several sub-bands.  The maximal DOS is found substantially larger than the DOS of the underlying 2DES.
The effect of the super-lattice potential to the SC gap $\Delta_0$ is examined by using BCS mean field theory.
Our theory is expected to be relevant to SC ultra-thin film of Pb or In with few monolayer (ML) if the proposed molecular graphene could be fabricated on these films. Note that a thin film of Pb with only a few ML may be approximately described by a single sub-band of 2DES\cite{shih2009,choudft,qwsargue}.
In our numerical calculations below, we use parameters similar to the thin films of Pb or In for the purpose of illustration.
In Pb thin films, the effective mass is typically between $1\sim10$ $m_e$ with $m_e$ the free electron mass depending on the coupling with the substrates.
The effective mass and the electron density can be controlled by choosing proper fabrication conditions, i.e. the protective cover,
the wetting layer, and different substrates, etc\cite{dil2007,dil2011,jeng2010,chiang2008}.
We believe that some of the results obtained remain qualitatively the same.

\begin{figure}
\centering
\includegraphics[width=8cm]{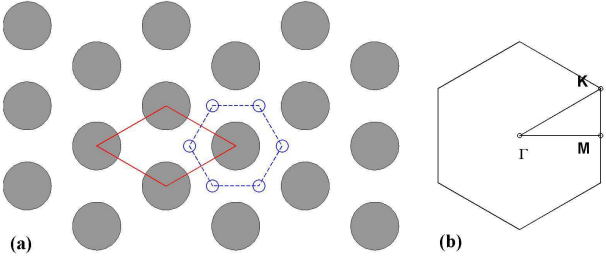}
\caption{(Color online). (a) A muffin potential with hexagonal lattice. The potential is $U_0>0$ inside the gray circles, and zero elsewhere.
Red lines indicate the unit cell of the lattice. Blue circles illustrate the sites of electrons in thin film. (b) The corresponding Brillouin zone.
}\label{fig1}
\end{figure}

We consider a 2DES with a hexagonal lateral potential lattice as illustrated in Fig. \ref{fig1}. The Hamiltonian is given by
\begin{eqnarray}
H=H_0 + H_{\textrm{pairing}},
\end{eqnarray}
with $H_{\textrm{pairing}}$ the electron pairing coupling [Eq. (5) below], and $H_0$ the single electron Hamiltonian for the normal state, which reads
\begin{eqnarray}
H_0 = H_{\textrm{2DES}}+\int dr V_{\textrm{muff}}(r) \rho(r),
\end{eqnarray}
where $H_{\textrm{2DES}}$ is  the free electron Hamiltonian of the underlying metal thin film given by
\begin{eqnarray}
H_{\textrm{2DES}}=\sum_{k\sigma} \frac{\hbar^2k^2}{2m^*} c^+_{k\sigma}c_{k\sigma} - \mu N,
\end{eqnarray}
with $k$ the wave-vector in 2D x-y plane, $m^*$ the effective mass, $\mu$ the chemical potential and N the total number of electrons.
The second part in $H_0$ describes the super-lattice potential, which is of a muffin tin type of hexagonal lattice with the lattice constant $a$.
The potential is $U_0 >0$ within the disks of a diameter $d$ and $0$ outside the disks as shown in Fig. 1(a).

The system described by $H_0$ has a discrete translational symmetry of a hexagonal lattice.
The supercell Brillouin zone (BZ) in the reciprocal lattice is illustrated in Fig. 1(b).
The single electron state of $H_0$ is described by a mini-band index $n$ and a wave-vector $k$ in the BZ,
and can be diagonalized with the plane wave basis, and we find,
 \begin{equation}
 H_{0}=\sum_{nk\sigma} \epsilon_{n\sigma}(k) c^+_{nk\sigma} c_{nk\sigma}
 \end{equation}
with $\epsilon_{n\sigma}(k)$ the energy dispersion.

\begin{figure}
\centering
\includegraphics[width=8.5cm]{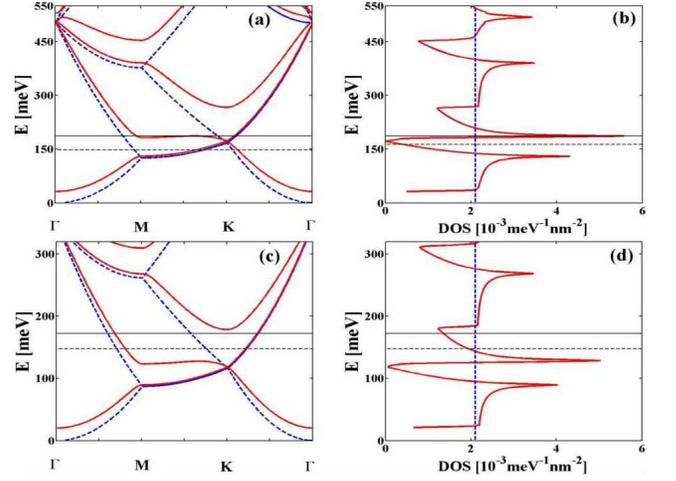}
\caption{(Color online). (a) Calculated energy bands (red solid lines) of the model at $a=2$ nm and $U_0=1$ eV.
The energy bands without super-lattice potential (namely $U_0=0$) are shown by blue dashed lines as a reference for comparison.
The Fermi level of the original 2DES ($U_0=0$, indicated by black dashed line) is lifted upward as indicated by the black slid line for $U_0=1$ eV.
(b) The corresponding density of state for $a=2$ nm case. (c) Energy bands (red solid lines) of the model at $a=2.4$ nm and $U_0=1$ eV
and the Fermi level (black solid line). (d) The corresponding density of state for $a=2.4$ nm case.
In the calculation, the number of electrons is fixed and the average electron density is $0.62\textrm{nm}^{-2}$ including spin degree of freedom.
Other parameters are $m^*=m_e$,  $d=0.5$ nm.}
\end{figure}

 In Fig. 2 (a) and (c), we plot the dispersion for a few lowest mini-bands along high symmetry directions in the BZ for parameters specified in the figure caption.
 The upper and bottom panels show the dispersions for the super-lattice constant $a$ to be 2 nm and 2.4 nm, respectively.
 The energy dispersion $E(k)$ in the lowest mini-band around the wave-vector $K$ is linear, demonstrating the massless Dirac fermion nature.
 The same dispersion occurs around wave-vector $K'$, which is not shown in this figure.  In Fig. 2(b) and (d), we plot the corresponding DOS or $N(E)$.
 As we can see,  $N(E(K))=0$ at the Dirac point, and $N(E) \propto |E(k) - E(K)|$, namely linear near $E(K)$, which reflects the Dirac fermion property.
 The DOS oscillates as a function of energy and several DOS peaks appears in contrast to the constant DOS of underlying 2DES.
 The oscillation is found to be very weak for higher index mini-band ($n=6$ or above).
 The position of Fermi level of the molecular graphene is related to the electron density $n_e$ (i.e.  chemical potential $\mu$) of the underlying 2DES
 if we assume that the fabrication process to generate the super-lattice potential does not alter the total number of electrons in the system.
 It aslo depends on the lattice constant $a$ and the diameter $d$ of the potential disk in Fig. 1.
 The Fermi levels at the zero temperature are indicated in Fig. 2 for a choice of the electron density $n_e=0.62 \textrm{nm}^{-2}$
 (including the spin degree of freedom), for $a= 2 nm$ and $a= 2.4 nm$, respectively.  The DOS at the Fermi level of the molecular graphene
 is near the maximum for $a=2 nm$ and is however much smaller  for $a= 2.4 nm$, as we can see from the figure.
 These results clearly show that the DOS at Fermi level can be modified by changing the super-lattice distance or the distance  between absorbed molecules.
 Note that  $U_0$ and $d$ determine the energy window of the Dirac fermion. If $U_0$ is too weak, or $d$ is too small,
 the Dirac point will overlap with other mini-bands and  there will be no linearity of the DOS.
\begin{figure}
\centering
\includegraphics[width=8.5cm]{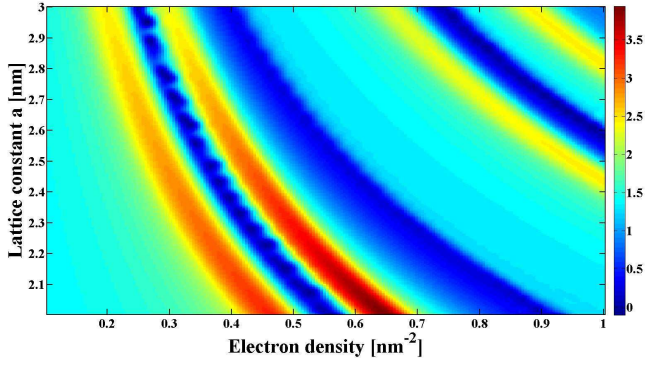}
\caption{(Color online). $\Delta_0$ (meV) in $a-n_e$ plane. The parameters are the same as the Fig. 2.}
\end{figure}

We now discuss superconductivity in this system by including a pairing Hamiltonian $H_{\textrm{pairing}}$.
\begin{equation}
H_{\textrm{pairing}}=  - \sum_{nn'kk'} V_{nk,n'k'}c^+_{nk\uparrow}c^+_{n\bar{k}\downarrow} c_{n'\bar{k'}\downarrow} c_{n'k' \uparrow}
\end{equation}
where $\bar{k}=-k$. The pairing interaction $V_{nk,n'k'} = V$ if the energies of the states $|nk\rangle$
and $|n'k'\rangle$ are both in the window of $(\mu-\hbar \omega_D, \mu+\hbar \omega_D)$, and $V_{nk,n'k'}=0$ otherwise, with $\hbar \omega_D$ the Debye energy cutoff.
We use BCS mean field theory to solve $H$ and the mean field Hamiltonian is given by
\begin{eqnarray}
H_{\textrm{MF}}=H_0- \sum_{nk} (\Delta c^+_{nk\uparrow}c^+_{n\bar{k}\downarrow} + h.c. )
\end{eqnarray}
where the SC gap $\Delta = \sum_{n'k'} V_{nk,n'k'} \langle c_{n'\bar{k'} \downarrow} c_{n'k' \uparrow}\rangle$ is solved by using the gap equation self-consistently
 \begin{eqnarray}\label{gap}
 \frac{1}{V}=\frac{1}{2}\sum_{|\epsilon_n(k)|<\hbar \omega_D} \frac{\textrm{tanh}(\frac{\beta E_{nk}}{2} )}{E_{nk}},\\
 \langle N \rangle = \sum_{nk} [1 - \frac{\epsilon_n (k)}{E_{nk}}\textrm{tanh}( \frac{\beta E_{nk}}{2} ) ]
 \end{eqnarray}
 where $E_{nk} = \sqrt{\epsilon^2_n (k) + \Delta^2}$ is the energy of the Bogoliubov quasi-particle. In the calculations below we use $\hbar \omega_D = 8.27 \textrm{meV}$,
 and $V= 180.54 \textrm{meV} \cdot \textrm{nm}^2$\cite{coupling}. These parameters may be suitable or abstracted from the known data on ultrathin film Pb.
 The results for the BCS gap at zero temperature $\Delta_0$ are shown in Fig. 3. We plot $\Delta_0$ in $a-n_e$ plane in Fig. 3, which is our main results.
 The SC gap oscillates with the artificial lattice constant $a$ or the electron density $n_e$. $\Delta_0$ vs $n_e$ for several choices of $a$ are shown in Fig. 4 (a).
 The maximum of the SC gap is 3.92 meV. This is significantly larger than the estimated SC gap of 1.08 meV for the system with the same set of parameters
 without the lateral potential (i.e. the SC gap of 4ML Pb thin film\cite{shih2006,shih2009}).  In passing, we note that when the chemical potential lies at the Dirac point,
 the DOS is zero, and there exists a threshold for the coupling $V$ in the Dirac fermion system, below which the gap is zero\cite{ezhao2006,jyuan2012}.
 Our result from the calculation is consistent with this argument. In Fig. 4(b), we show $\Delta_0$ vs $a$ for three values of electron density.
  In experiment, the molecule super-lattice constant a is controllable and the electron density can be adjusted by gating.
  Thus, the two kinds of SC gap oscillation are observable.\begin{figure}
\centering
\includegraphics[width=8.5cm]{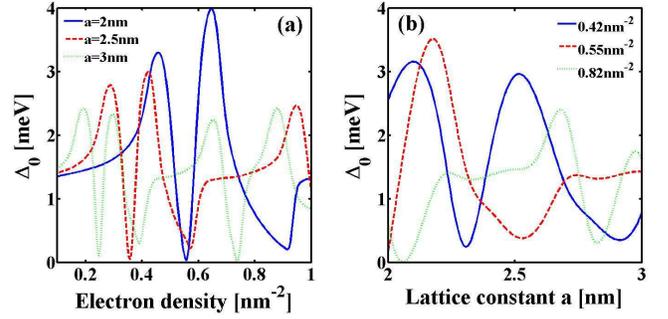}
\caption{(Color online). (a) $\Delta_0$ as a function of $n_e$ for given lattice constant a. (b) $\Delta_0$ as a function of lattice constant a for given $n_e$. Other parameters are the same as Fig. 2}
\end{figure}
   It is interesting to note that the change of the SC gap by tuning the parameters of the molecule lattice is dramatic,
   due to the dramatic change of the DOS at the Fermi level, thanks to the Van Hove singularities of the DOS.
   Note that the BCS mean field theory applies to weak coupling case. In the strong pairing interaction,
   the above estimate is not accurate and more sophisticated theory will be needed for quantitative estimate.
   For instance, Pb is known to be a strong coupling superconductor.  Nevertheless, the dramatic change in the DOS is expected to
   lead to dramatic changes in the SC gap.



In summary,  we have proposed to fabricate a molecule super-lattice  on a SC thin film to manipulate superconductivity of 2D superconductor.
We study the superconductivity in a molecule graphene system, which has been experimentally realized recently.  In molecule graphene, a lateral super-lattice is applied on a 2DES by molecule absorbing, and the dispersion becomes linear in analogy to the case of graphene.
The density of states at the  Fermi level as well as the superconducting gap of the system may be dramatically modified
by adjusting the molecule lattice. With the Pb ultrathin film as an example,
our calculations show that it is an efficient way to manipulate the 2D superconductivity.
The predicted changes of the superconducting gap can be detected by STM experiment.
Assuming the 2D superconductivity is stabilized by the quasi-2D nature and our calculation implies the superconducting transition temperature
may be increased dramatically by the molecule super-lattice.
The important issue is that this work points out that nano-fabrication is a promising way to control  superconductivity in 2D system.
Nowadays,  due to the rapid progress of nanotechnology, various techniques have been developed to make a desired molecule lattice on metal thin film\cite{self1,self2}.
Meanwhile,  many novel 2D superconductors have been reported in experiments, including one unit cell FeSe on SrTiO3\cite{fese},
two-atomic-layer crystalline Ga film\cite{ga},  the electric field induced superconductivity on the surface of $\textrm{SrTiO}_3$\cite{ueno},
and 2D $MoS_2$\cite{ye}.
Thus, we believe our proposal could be realized in future experiments. Furthermore, in addition to the hexagonal molecule super-lattice that we studied in this paper,
other kinds of super-lattice, though  not been reported in experiment so far, may also influence DOS as well as SC gap.

J.H.G acknowledges support from the National Natural Science Foundation of China (Grants No. 11274129).
YZ is supported by National Basic Research Program of China (No.2011CBA00103/2014CB921201),
NSFC (No.11374256) and the Fundamental Research Funds for the Central Universities in China.
FCZ is supported by National Basic Research Program of China (No.2014CB921203) and NSFC (No.11274269).

\end{document}